\documentstyle[12pt]{article}
\newcommand{\thspace}{\kern.08333em}

\def \beq{\begin{equation}}
\def \eeq{\end{equation}}
\def \beqn{\begin{eqnarray}}
\def \eeqn{\end{eqnarray}}

\def \Abar{\bar A}
\def \Dbar{\bar{D}^0}

\def \ubar{\bar {u}}
\def \cbar{\bar {c}}
\def \sbar{\bar{s}}
\def \s{\sqrt{2}}

\def \beq{\begin{equation}}
\def \eeq{\end{equation}}
\def \beqn{\begin{eqnarray}}
\def \eeqn{\end{eqnarray}}

\def \b{{\cal B}}
\def \Dbar{\bar{D}^0}

\def \ubar{\bar {u}}
\def \cbar{\bar {c}}
\def \sbar{\bar{s}}
\def \s{\sqrt{2}}

\begin{document}
\rightline{CALT-68-2159}
\rightline{hep-ph/9802XXX}
\rightline{February 1998}
\bigskip
\bigskip
\centerline{\bf WEAK PHASE $\gamma$ FROM COLOR-ALLOWED $B \to D K$ RATES}
\bigskip
\centerline{\it Michael Gronau\footnote{Permanent Address: Physics Department,
Technion -- Israel Institute of Technology, 32000 Haifa, Israel.}} 
\centerline{\it California Institute of Technology}
\centerline{\it Pasadena, CA 91125}
\bigskip
\centerline{\bf ABSTRACT}
\vskip 2cm  

\begin{quote}
The ratios of partial rates for charged $B$ decays to the recently 
observed $D^0 K$ mode and to the two $D_{CP}K$ final states (CP = $\pm$)
are shown to constrain the weak phase $\gamma \equiv {\rm Arg} (V_{ub}^*)$. 
The smaller color-suppressed rate, providing further information about the 
phase, can be determined from these rates alone. Present estimates 
suggest that, while the first constraints can already be obtained in a
high luminosity $e^+e^-$ $B$ factory, measuring the color-suppressed rate 
would require dedicated hadronic $B$ production experiments.  
\end{quote}

\vskip 3cm

\leftline{\qquad PACS codes:  12.15.Hh, 12.15.Ji, 13.25.Hw, 14.40.Nd}
\vfill
\newpage

The CLEO Collaboration has recently observed the decay $B^-\to
D^0 K^-$ and its charge conjugate \cite{CLEO}. This is the first 
observation of a decay mode described by the                         
quark subprocess $b \to c \ubar s$ involving the Cabibbo-Kobayashi-Maskawa 
(CKM) factor $V_{cb}V^*_{us}$. The reported branching ratio, $0.055 \pm 
0.014 \pm 0.005$, measured relative to $B^- \to D^0 \pi^-$, is in agreement 
with the Standard Model expectation. The observed decay plays an crucial 
role in a method proposed some time ago \cite{GW,GL} to determine the CP 
violating weak phase $\gamma$, the relative phase between 
$V_{cb}V^*_{us}$ and $V_{ub}V^*_{cs}$. The purpose of this Letter is 
to reexamine this method in view of its importance, and to suggest some 
variants to overcome its difficulties. A complementary variant was proposed in 
Ref.~\cite{ADS}.

The other processes involved in the method are $B^- \to D_{CP} K^-,~B^- \to 
\Dbar K^-$ and their charge conjugates. Partial $D$ 
decay rates into CP-eigenstates (such as $K^+K^-$) are about 
an-order-of-magnitude smaller than into states of specific flavor (such as 
$K^-\pi^+$). Thus, by combining a few CP modes, the decays 
$B^- \to D_{CP} K^-$ should be observed in near future high statistics 
experiments. The third process, $B^- \to \Dbar K^-$, mediated by $b \to u \cbar s$ and involving the CKM factor $V_{ub}V^*_{cs}$, is harder to measure. 
It is usually assumed to have a ``color-suppressed" branching ratio, 
about two orders-of-magnitude smaller than that of $B^-\to D^0 K^-$. Let us 
recall the arguments on which this estimate is based.

The effective Hamiltonians 
for $b \to c \ubar s$ and $b \to u \cbar s$ transitions are
\beq
H_{\rm eff}(b\to c \ubar s) = \frac{G_F}{\s} V_{cb}V^*_{us}[c_1(\mu)(\sbar u)
(\cbar b) + c_2(\mu)(\cbar u)(\sbar b)]~~,
\eeq
and
\beq
H_{\rm eff}(b\to u \cbar s) = \frac{G_F}{\s} V_{ub}V^*_{cs}[c_1(\mu)(\sbar c)
(\ubar  b) + c_2(\mu)(\ubar c)(\sbar b)]~~,
\eeq
respectively, where $c_1(m_b)=1.13,~c_2(m_b)=-0.29$ \cite{BURAS}. $(\cbar
b)=\cbar \gamma^{\mu}(1-\gamma_5)c$ etc. are left-handed color-singlet quark
currents.  The ratio of the corresponding CKM factors is
$|V_{ub}V^*_{cs}/V_{cb}V^*_{us}| = 0.4 \pm 0.1$ \cite{DRELL}. The hadronic
matrix elements of the four-fermion operators, depending on the scale $\mu$,
are very difficult to calculate. The conventional description of
strangeness-conserving decays such as $B^0 \to D^- \pi^+$ assumes that
``color-allowed" operator matrix elements factorize \cite{BSW}. Nonperturbative
effects, arising from soft gluon exchange \cite{DGS}, require the use of a free
parameter to describe decay amplitudes. This parameter, fitted by data,
determines the ratio of color-suppressed and color-allowed amplitudes, $a_2/a_1
\approx 0.26$ \cite{BROWDER}. This value depends on unmeasured form factors of
$B$ mesons into light mesons for which a model must be assumed . These form
factors dominate color-suppressed amplitudes of processes such as $B^0 \to
\Dbar \pi^0$. Using flavor SU(3) \cite{GHLR}, this value of $a_2/a_1$ can also
be used to study $B \to \bar{D} K$ decays. An application to relations between
$B \to \bar{D} K$ (given by a $B \to D$ form factor) and $B \to D K$ 
(given by a $B \to K$ form factor), in which final states carry
opposite charm, is more questionable. Nevertheless, one often assumes that 
\beq\label{r}
r \equiv \frac{|A(B^- \to \Dbar K^-)|}{|A(B^- \to D^0 K^-)|} \approx 
\frac{|V_{ub}V^*_{cs}|} {|V_{cb}V^*_{us}|} \frac{a_2}{a_1} \approx ~~0.1~~.
\eeq

It is difficult to associate a theoretical uncertainty with this estimate,
which is based largely on empirical observations in the $\Delta S = 0$ sector,
rather than on firm theoretical grounds.
We will usually assume that the ratio of amplitudes $r$ cannot be greater or 
smaller than 0.1 by a factor larger than two. We note, however, that larger 
values cannot be excluded. As we will see, the precision
of determining the weak phase $\gamma$ improves as $r$ increases. One of the 
questions addressed in the present Letter is how to determine this quantity 
experimentaly.

An essential difficulty in measuring the rate of $B^- \to \Dbar K^-$ was 
pointed out by Atwood, Dunietz and Soni \cite{ADS}.
When a $\Dbar$ from $B^- \to \Dbar K^-$ is identified through its hadronic 
decay mode (such as $K^+\pi^-$), the decay amplitude interferes with a 
comparable doubly Cabibbo suppressed decay amplitude of a $D^0$ from 
$B^- \to D^0 K^-$. (Here Eq.~(\ref{r}) is assumed).
This forbids a direct measurement of 
$\Gamma(B^-\to\Dbar K^-$). Using two different final states to identify a 
neutral $D$ meson, (e.g. $K^+\pi^-$ and $K^+\pi^-\pi^0$), may 
allow a determination of $r$ and $\gamma$ from the branching ratios of these 
processes and their charge-conjugates \cite{ADS}. The  
products of corresponding $B$ and $D$ decay branching ratios are expected to be about two orders of magnitude smaller than $B(B^-\to D^0 K^-)B(D^0 \to K^- 
\pi^+)$, at a level of $10^{-7}$. The number of events, expected in future 
$e^+e^-$ colliders, is likely to be too small to allow a precise determination 
of $r$ and $\gamma$ \cite{SOFFER}. Such precision can potentially be achieved 
in dedicated hadronic $B$ production experiments \cite{STONE}, which are 
expected to yield an order of a few thousand events of this kind \cite{STONE2}.

Let us study the information about $\gamma$ obtained 
from measuring only the more abundant processes $B^-\to D^0 K^-,~ B^- \to 
D_{CP} K^-$ and their charge-conjugates. 
We will derive a simple sum rule from which the suppressed rate of 
$B^-\to\Dbar K^-$ can, in principle, be determined from the above less 
suppressed rates, without involving an interference with $B^- \to D^0 K^-$.
New constraints on the weak phase $\gamma$ will be shown to be obtained 
by measuring only the two ratios of partial decay rates into CP-even-and-odd
and into flavor states, combining particles and antiparticles. We will 
look into the prospects of carrying out these studies in a future very high 
luminosity $e^+e^-$ $B$ factory. 

Defining decay amplitudes by their magnitudes, strong and weak phases,
\beq
A(B^+ \to \Dbar K^+) = \Abar e^{i\bar{\Delta}}~~,~~~~~
A(B^+ \to D^0 K^+) = A e^{i\Delta} e^{i\gamma}~~,
\eeq
we find (disregarding common phase space factors)
\beq
\Gamma(B^+ \to \Dbar K^+) = \Gamma(B^- \to D^0 K^-) = \Abar^2~~,
\eeq
\beq
\Gamma(B^+ \to D^0 K^+) = \Gamma(B^- \to \Dbar K^-) = A^2~~,
\eeq
\beq
\Gamma(B^{\pm} \to D_1 K^{\pm}) = \frac{1}{2}[\Abar^2 + A^2 + 2\Abar A \cos 
(\delta \pm \gamma)]~~,
\eeq
\beq
\Gamma(B^{\pm} \to D_2 K^{\pm}) = \frac{1}{2}[\Abar^2 + A^2 - 2\Abar A \cos 
(\delta \pm \gamma)]~~,
\eeq
where $\delta \equiv \Delta-\bar{\Delta}$.~~$D_{1,2}$ are the two neutral $D$ 
meson CP-eigenstates, $D_{1,2}=(D^0 \pm \Dbar)/\s$.

One obtains the following sum rule
\beq                       
\Gamma(B^- \to D_1 K^-) + \Gamma(B^- \to D_2 K^-) =
\Gamma(B^- \to D^0 K^-) + \Gamma(B^- \to \Dbar K^-)~~.
\eeq
A similar sum rule is obeyed by the charge-conjugated processes.
In principle these sum rules allow a determination of $\Gamma(B^- \to 
\Dbar K^-)=\Gamma(B^+ \to D^0 K^+)$ from measurements of the other larger rates.
Using Eq.~(\ref{r}) we note, however, that the second rate on the 
right-hand-side is 
expected to be about two orders-of-magnitude smaller than the first rate. 
Therefore, a useful determination of  $\Gamma(B^- \to \Dbar K^-)$ requires  
very precise measurements of $\Gamma(B^- \to D^0 K^-)$ and of $\Gamma(B^- 
\to D_{CP} K^-)$.
    
The ratio of amplitudes $r$ can be obtained from the charge-averaged ratio for 
decays into $D$ meson CP and flavor states
$$
S \equiv \frac{\Gamma(B^+ \to D_1 K^+) +\Gamma(B^- \to D_1 K^-) + 
\Gamma(B^+ \to D_2 K^+) +\Gamma(B^- \to  D_2 K^-)}
{\Gamma(B^+ \to \Dbar K^+) +\Gamma(B^- \to  D^0 K^-)}~~,
$$
\beq\label{S}
S = 1 + r^2~~.
\eeq
The CP asymmetries of decays into $D_1 K$ and $D_2 K$, normalized by the rate
into the $D$ meson flavor state, 
\beq\label{A_i}
{\cal A}_i \equiv \frac{\Gamma(B^+ \to D_i K^+) - \Gamma(B^- \to D_i K^-)}
{\Gamma(B^+ \to \Dbar K^+) + \Gamma(B^- \to D^0 K^-)}~~,~~~~~i = 1,2~,
\eeq
are equal in magnitude and have opposite signs. They yield a combined asymmetry
\beq
{\cal A} \equiv {\cal A}_2 - {\cal A}_1 = 2r \sin\delta \sin\gamma~~.
\eeq 

It is convenient to define two charge-averaged ratios for the
two CP-eigenstates
\beq\label{R_i}
R_i \equiv \frac{2[\Gamma(B^+ \to D_i K^+) + \Gamma(B^- \to D_i K^-)]}
{\Gamma(B^+ \to \Dbar K^+) + \Gamma(B^- \to D^0 K^-)}~~,~~~~~i = 1,2~~,
\eeq
for which we find
\beq\label{R12}
R_{1,2} = 1 + r^2 \pm 2r\cos\delta\cos\gamma~~.
\eeq
The factor of 2 in the numerator of $R_{1,2}$ is used to
normalize these ratios to values of approximately one.
Rewriting
\beq
R_{1,2} = \sin^2\gamma + (r \pm \cos\delta\cos\gamma)^2 + 
\sin^2\delta\cos^2\gamma~~,
\eeq
one obtains the two inequalities
\beq\label{LIMIT}
\sin^2\gamma \leq R_{1,2}~~,~~~~~i = 1,2~~.
\eeq
                     
The quantities $S,~{\cal A}$ and $R_i$ hold information from which $r,~\delta$ and
$\gamma$ can be determined up to discrete ambiguities. 
$r$ is given by $S$, and $\gamma$ is obtained from $R_i$ and ${\cal A}$:
\beq
R_i = 1 + r^2 \pm \sqrt{4 r^2 \cos^2 \gamma - {\cal A}^2 \cot^2 \gamma}~~.
\eeq
Plots of $R_i$ as function of $\gamma$ for a few values of $r$ and ${\cal A}$,
and the precision in $r,~R_i$ and ${\cal A}$ required to measure $\gamma$ to a 
given level, are given in Ref.~\cite{KPI}.   
The accuracy of this method of determining $\gamma$ depends on the actual 
value of $r$. The larger this ratio, the more precisely can $\gamma$ be 
determined. 

If $r$ is as small as estimated in Eq.~(\ref{r}), then a useful determination 
of this quantity from $S$ requires that the rates in the numerator 
and denominator of Eq.~(\ref{S}) are measured to better than 1$\%$
which is unattainable in near future experiments.
This demonstrates the difficulty of looking for the color-suppressed process.

Consider, for instance, a sample of 300 million $B^+B^-$ pairs thought to be 
produced in an upgraded version of CESR \cite{WEINSTEIN}. Using $\b(B^- \to D^0
K^-) = 3 \times 10^{-4}$ \cite{CLEO} and $\b(D^0\to K^-\pi^+)=0.04$ \cite{PDG}, 
one expects a total of about 7000 identified $D^0 K^-$ and $\Dbar K^+$ events.
(Use of other $D$ decay modes compensates for suppression due to detection 
efficiencies). This would yield a
1.2$\%$ measurement of the sum of rates in the denominator of Eq.~(\ref{S}). To
estimate the precision of the numerator, in which the $D$ meson decays to
CP-eigenstates, we use \cite{PDG} $\b(D^0 \to \pi^+\pi^-) + \b(D^0 \to K^+ K^-)
= 6\times 10^{-3}$ for two positive CP states. This implies a combined sample
of about 1000 identified $D_1 K^+$ and $D_1 K^-$ events. (Detection 
efficiencies may decrease this number somewhat). Let us assume a
similar number of $D_2 K^+$ and $D_2 K^-$ events, identified by $D$ decay final
states such as $K_S \pi^0,~K_S\rho^0$ and $K_S \phi$. (The combined decay 
branching ratio
into these states and others are actually larger than into positive CP-eigenstates \cite{PDG}, however detection efficiencies are smaller due to the larger 
number of final particles). This determines the numerator to within 
2.2$\%$, so that the combined statistical error on $S$ is 2.5$\%$. Systematic 
uncertainties are likely to increase this error, although some of them cancel
in the ratio of rates. Assuming that the total error in $S$ is 5$\%$, a
90$\%$ c.l. upper limit $r < 0.25$ could be obtained from this measurement.

Another way for learning $r$ is by measuring 
the rate of the rare process $B^- \to (K^+\pi^-)_D K^-$ combined with its 
charge-conjugate. 300 million $B^+B^-$ pairs lead to a few tens of events,
for which a large error in the combined branching ratio is expected.  
The amplitude of this process consists of 
two interfering contributions carrying an unknown relative phase. The two 
terms describe the color suppressed process $B^- \to \Dbar K^-$ followed by 
Cabibbo-allowed $D$ decay, and $B^- \to D^0 K^-$ followed by a doubly Cabibbo 
suppressed $D$ decay. The magnitude of the second amplitude is expected to be 
known to a few percent at the time of the experiment. Comparison of  this 
amplitude with the measured one could be used to constrain $r$. In the likely
case that the two amplitudes are equal within experimental errors, 
destructive interference
would be assumed to obtain an upper limit on $r$,~~$r < 2\sqrt{\b(D^0 \to 
K^+\pi^-)/\b(\Dbar \to K^+\pi^-)} = 0.18$.  Here current central values were 
used and all experimental errors were neglected. This upper limit, increased 
somewhat by experimental errors in branching ratios, is about the limit 
obtained from $S$.

Assuming that $r$ is too small to be measured from $S$ 
(i.e. $r \leq 0.2$), one may still obtain useful constraints on the weak 
phase $\gamma$ from the asymmetries ${\cal A}_i$ and the two ratios $R_i$. The 
information obtained from these pairs of quantities is complementary to 
each other. While the asymmetries become larger for large values of 
$\sin\delta\sin\gamma$, the deviation of $R_i$ from $1 + r^2 \approx 1$ 
increases with $\cos\delta \cos\gamma$. One thousand
identified $D_iK^{\pm}$ events allow a $3\sigma$ asymmetry measurement at a
level of 10$\%$ or larger. For $r\sim 0.1$, such asymmetries are expected
if $\delta$ is sizable, namely $\delta > 30^{\circ}$. It is needless to 
emphasize the importance of nonzero CP asymmetry measurements, however one
should foresee a possibility of small final state phases. Upper limits on 
the corresponding final state phase-difference in $B \to \bar{D} \pi$ decays 
are already at a level of $20^{\circ}$ \cite{NELSON}. Assuming that the final 
state phase-difference between $B \to D K$ and $B \to \bar{D} K$ is not larger, the only information about $\gamma$ would be derived from $R_i$. 

A particularly interesting case is $R_i < 1$, holding for either
$i=1$ or $i=2$. Using Eq.~(\ref{LIMIT}), this implies
new bounds on $\gamma$. The condition $R_i < 1$ ($i=1$~or~$2$), equivalent to
$|\cos \delta \cos \gamma| > r/2$, holds for
values of $\delta$ and $\gamma$ which are not too close to 90 degrees.
Taking $\cos \delta \approx 1$ and $r\sim 0.1$, this condition is fulfilled by 
all the currently allowed vallues of $\gamma$, $30^{\circ} \leq \gamma \leq
150^{\circ}$ \cite{GAMMA}, excluding a narrow band around $\gamma=90^{\circ}$.
That is, {\it for all values outside this narrow band one of the two ratios of
rates $R_1$ or $R_2$ must be smaller than one}, $R_1 < 1$ for $\gamma > 
90^{\circ}$ and $R_2 < 1$ for $\gamma < 90^{\circ}$.

Using $r=0.1,~\delta=0$ in
Eq.~(\ref{R12}), we find for $\gamma = 150~(30),~140~(40),~130~(50),
120~(60)$, $110~(70),~100~(80)$ degrees the following values of $R_1~(R_2)$:
$0.84,~0.86,~0.88,~0.91,~0.94$, $0.98$, respectively. Measuring these values 
for $R_1$
or $R_2$ would exclude by Eq.~(\ref{LIMIT}) the following ranges of $\gamma$:
$66^{\circ}-114^{\circ},~68^{\circ}-112^{\circ},~70^{\circ}-110^{\circ},
73^{\circ}-107^{\circ},~76^{\circ}-104^{\circ},~81^{\circ}-99^{\circ}$,
respectively. For another choice of parameters, $r=0.2,~ \delta=20^{\circ}$,
the measurements of $R_i$ corresponding to the above values of $\gamma$ would
be 0.71, 0.75, 0.80, 0.85, 0.91, 0.97, respectively. These values exclude 
somewhat
larger ranges of $\gamma$ than in the case $r=0.1$. Assuming the above number
of events, the statistical error of measuring $R_1$ and $R_2$ is 3.4$\%$. 
Particularly interesting is the ratio of rates $R_1$. The systematic errors in 
this ratio, in which the numerator and denominator
involve similar three charged pion and kaon final states, are expected to 
cancel. A few percent accuracy in $R_i$ is sufficient for excluding a sizable 
range of 
values of $\gamma$ for the above two choices of parameters. The excluded range 
grows with increasing values of $r$, for which smaller values of $R_i$ are 
obtained. 

In conclusion, we have shown that the ratios of rates $R_i$, for charged $B$ 
decays to the two $D_{CP}K$ final states and to the already observed $D^0 K$ 
mode, lead to new constraints on the weak phase $\gamma$.
The smaller color-suppressed rate, which would lend further information 
about this phase, can  be determined from a sum rule involving 
these rates. The estimate of Eq.~(\ref{r}) suggests that this may be beyond the 
capability of future $e^+e^-$ $B$ factories, and would have to await
dedicated hadronic $B$ production experiments. This method is complementary to 
the one suggested in Ref.~\cite{ADS}. The two methods seem to be comparable in 
their statistical power for
determining $\gamma$ from the color-suppressed rate of $B^- \to \Dbar K^-$, 
which requires in both cases very high statistics hadronicly produced $B$ 
experiments. With less statistics, already available in high luminosity 
$e^+e^-$ experiments, the present method can be used to set new bounds on 
$\gamma$ through measurements of the more abundant processes 
$B^{\pm} \to D_{CP} K^{\pm}$.

\bigskip

I wish to thank J. Rosner, A. Soffer, S. Stone, M. Wise and F. W\"urthwein for 
useful discussions.  I am grateful to the Caltech High Energy Theory Group for 
its kind hospitality. This work was supported by the United States Department 
of Energy under Grant No.~DE-FG03-92-ER40701. 
\bigskip

\def \ajp#1#2#3{Am.~J.~Phys.~{\bf#1}, #2 (#3)}
\def \apny#1#2#3{Ann.~Phys.~(N.Y.) {\bf#1}, #2 (#3)}
\def \app#1#2#3{Acta Phys.~Polonica {\bf#1}, #2 (#3)}
\def \arnps#1#2#3{Ann.~Rev.~Nucl.~Part.~Sci.~{\bf#1}, #2 (#3)}
\def \cmp#1#2#3{Commun.~Math.~Phys.~{\bf#1}, #2 (#3)}
\def \ib{{\it ibid.}~}
\def \ibj#1#2#3{~{\bf#1}, #2 (#3)}
\def \ijmpa#1#2#3{Int.~J.~Mod.~Phys.~A {\bf#1}, #2 (#3)}
\def \ite{{\it et al.}}
\def \jmp#1#2#3{J.~Math.~Phys.~{\bf#1}, #2 (#3)}
\def \jpg#1#2#3{J.~Phys.~G {\bf#1}, #2 (#3)}
\def \mpla#1#2#3{Mod.~Phys.~Lett.~A {\bf#1}, #2 (#3)}\def \ib{{\it ibid.}~}
\def \ibj#1#2#3{~{\bf#1}, #2 (#3)}
\def \ijmpa#1#2#3{Int.~J.~Mod.~Phys.~A {\bf#1}, #2 (#3)}
\def \ite{{\it et al.}}
\def \jmp#1#2#3{J.~Math.~Phys.~{\bf#1}, #2 (#3)}
\def \jpg#1#2#3{J.~Phys.~G {\bf#1}, #2 (#3)}
\def \mpla#1#2#3{Mod.~Phys.~Lett.~A {\bf#1}, #2 (#3)}
\def \nc#1#2#3{Nuovo Cim.~{\bf#1}, #2 (#3)}
\def \npb#1#2#3{Nucl.~Phys. B~{\bf#1}, #2 (#3)}
\def \pisma#1#2#3#4{Pis'ma Zh.~Eksp.~Teor.~Fiz.~{\bf#1}, #2 (#3) [JETP
Lett. {\bf#1}, #4 (#3)]}
\def \pl#1#2#3{Phys.~Lett.~{\bf#1}, #2 (#3)}
\def \plb#1#2#3{Phys.~Lett.~B {\bf#1}, #2 (#3)}
\def \pr#1#2#3{Phys.~Rev.~{\bf#1}, #2 (#3)}
\def \pra#1#2#3{Phys.~Rev.~A {\bf#1}, #2 (#3)}
\def \prd#1#2#3{Phys.~Rev.~D {\bf#1}, #2 (#3)}
\def \prl#1#2#3{Phys.~Rev.~Lett.~{\bf#1}, #2 (#3)}
\def \prp#1#2#3{Phys.~Rep.~{\bf#1}, #2 (#3)}
\def \ptp#1#2#3{Prog.~Theor.~Phys.~{\bf#1}, #2 (#3)}
\def \rmp#1#2#3{Rev.~Mod.~Phys.~{\bf#1}, #2 (#3)}
\def \rp#1{~~~~~\ldots\ldots{\rm rp~}{#1}~~~~~}
\def \stone{{\it B Decays}, edited by S. Stone (World Scientific,
Singapore, 1994)}
\def \yaf#1#2#3#4{Yad.~Fiz.~{\bf#1}, #2 (#3) [Sov.~J.~Nucl.~Phys.~{\bf #1},
#4 (#3)]}
\def \zhetf#1#2#3#4#5#6{Zh.~Eksp.~Teor.~Fiz.~{\bf #1}, #2 (#3) [Sov.~Phys.
- JETP {\bf #4}, #5 (#6)]}
\def \zpc#1#2#3{Zeit.~Phys.~C {\bf#1}, #2 (#3)}

\end{document}